\documentclass[usenatbib]{article}
\usepackage{times,graphicx,natbib,a4wide, hyperref}

\usepackage{graphicx}	% Including figure files
\usepackage{amsmath}	% Advanced maths commands
\usepackage{amssymb}	% Extra maths symbols
\usepackage{pdflscape}	% Extra maths symbols
\usepackage[english]{babel}
\usepackage{multicol}
\usepackage{xcolor}
\newcommand{\degrees}{^{\circ}}
\newcommand{\msol}{M_{\rm \odot}}
\newcommand{\lsol}{L_{\rm \odot}}
\newcommand{\mjup}{M_{\rm Jup}}
\newcommand{\rjup}{R_{\rm Jup}}

\newcommand{\mearth}{M_{\rm \oplus}}
\newcommand{\rearth}{R_{\rm \oplus}}

\newcommand{\cotwo}{\rm{CO_2}}

\begin{document}

\title{The habitable zone for Earthlike exomoons orbiting Kepler-1625b}
\author{Duncan H. Forgan$^1$}
\maketitle

\noindent $^{1}$Centre for Exoplanet Science, SUPA, School of Physics \& Astronomy, University of St Andrews, St Andrews KY16 9SS, UK \\

\noindent \textbf{Word Count: 3,278} \\

\noindent \textbf{Direct Correspondence to:} \\
D.H. Forgan \\ \\
\textbf{Email:} dhf3@st-andrews.ac.uk \\
\textbf{Post:} Dr Duncan Forgan \\
Room 317, School of Physics and Astronomy
University of St Andrews \\
North Haugh, St Andrews \\
KY16 9SS, UK

\newpage

\begin{abstract}

The recent announcement of a Neptune-sized exomoon candidate orbiting the Jupiter-sized object Kepler-1625b has forced us to rethink our assumptions regarding both exomoons and their host exoplanets.  In this paper I describe calculations of the habitable zone for Earthlike exomoons in orbit of Kepler-1625b under a variety of assumptions.  I find that the candidate exomoon, Kepler-1625b-i, does not currently reside within the exomoon habitable zone, but may have done so when Kepler-1625 occupied the main sequence.  If it were to possess its own moon (a ``moon-moon'') that was Earthlike, this could potentially have been a habitable world.  If other exomoons orbit Kepler-1625b, then there are a range of possible semimajor axes/eccentricities that would permit a habitable surface during the main sequence phase, while remaining dynamically stable under the perturbations of Kepler-1625b-i.  This is however contingent on effective atmospheric $\cotwo$ regulation.

\end{abstract}


\section{Introduction}

\noindent Almost since the first detection of extrasolar planets (exoplanets, \citealt{Mayor1995}),  extrasolar moons (exomoons) have been the subject of intense scientific inquiry.  In the Solar System, moons act as tracers of planet formation and evolution.  In some cases, moons such as Europa, Enceladus and Ganymede may even host subsurface habitats \citep{Melosh2004,Iess2014,Thomas2015a,Saur2015}.  It is quite possibly the case that subsurface-habitable moons could greatly outnumber habitable planets in the Milky Way \citep{Scharf2006,Heller2015}.  It is also quite possibly the case that exomoons are massive enough to host a substantial atmosphere like the Earth, and possess a similar biosphere.

Earth-like exomoons have been thought to be unlikely based on Solar System evidence.  The moons of the giant planets possess a satellite-to-planet mass ratio $\epsilon \lessapprox 10^{-4}$, which is consistent with models of moon formation in a circumplanetary disc \citep{Mosqueira2003, Mosqueira2003a,Ward2010,Canup2006}.  For a Jupiter-mass planet to host an Earth-like exomoon, this would require $\epsilon \approx 3 \times 10^{-3}$, i.e. at least an order of magnitude higher.

Recently, \citet{Teachey2017} described evidence pointing to a candidate exomoon in orbit of the Jupiter-sized transiting exoplanet Kepler-1625b.  The host star, Kepler-1625, is G type, of approximately one solar mass and has recently evolved off the main sequence \citep{Berger2018}.

The exomoon candidate was observed in three transits of Kepler-1625b over the four-year primary mission of the Kepler Space Telescope, out of a maximum of five transits (the orbital period of Kepler-1625b being 287 days).  Transit T2 appears to indicate the satellite performing an early ingress of the transit, with T4 showing the satellite in late egress.  Interestingly, T5 shows both early ingress and late egress, where the moon lags some 10 hours behind the planet in exiting the stellar disc.  

Amongst other measurements, this gave initial constraints on not only the projected separation of the two bodies, but also the orbital period of the moon - around $P_{ps} \sim 72$ hours, and a separation of around 19 planetary radii ($R_P$).  Hence by Kepler's third law, the barycentric mass for the planet moon component was derived to be around $17.6^{+2.1}_{-1.9}\, \mjup$.

The mass for Kepler-1625b was not well-constrained, and hence the exomoon candidate (Kepler-1625b-i) also has a poorly constrained mass\footnote{It is also worth noting the sensitivity of these inferences to the detrending algorithm used \citep{Rodenbeck2018}}. \citet{Teachey2017} suggested that the planet is 10 $\mjup$, with the moon being around 17 $\mearth$, giving $\epsilon=5.34 \times 10^{-3}$.  \citet{Heller2017c} notes that the barycentric mass could be shared differently than \citet{Teachey2017}'s description.  Indeed, Kepler-1625b may be a brown dwarf or a low mass star, and the exomoon candidate mass would be closer to 1$\mearth$.

Subsequent observation with the Hubble Space Telescope (HST) found further evidence in favour of the exomoon's presence  \citep{Teachey2018}, favouring the 10$\mjup$ planet, Neptune-mass exomoon interpretation.  However, this inference is now dominated by the single HST observation, and the authors advocate further monitoring to confirm the moon-like signal.

In any case, if this candidate is confirmed by subsequent observations, this would indicate that satellite systems with $\epsilon \sim 10^{-3}$ can indeed exist, and that massive exomoons with Earthlike properties are indeed possible.

Massive exomoons raise the stakes for habitability even further.  One dimensional climate modelling of Earthlike exomoons in orbit of giant planets around Sunlike stars show that the habitable zone for such objects occupies a significant volume of orbital parameter space \citep{Forgan_moon1,Heller2013}.  This parameter space is generally larger in volume than that of Earthlike exoplanets due to the additional sources and sinks of radiation.

Planetary illumination allows the habitable zone to move further away from the host star, especially if the planet is sufficiently large that it is significantly self-luminous in thermal radiation \citep{Heller2013}.  Typically, the thermal flux dominates over reflected starlight, suggesting a slowly varying illumination, but it is worth noting that moons on eccentric orbits will still experience strong variations in illumination \citep{Hinkel2013}.

On the other hand, moons orbiting close to their host planet are more likely to experience relatively long eclipses of the star by the planet, which can result in a net loss of radiation \citep{Heller2012}.  While the loss from a single eclipse can usually be buffered by the thermal inertia of an Earthlike atmosphere, if the eclipses are frequent and sufficiently long in duration, moons can enter a snowball state which they cannot exit even after leaving the eclipse \citep{Forgan2014a}.

Tidal heating can allow the moon's habitability to be almost independent of the star, but by the same token can be strong enough to render the moon uninhabitable.  The complex nature of tidal heating demands careful modelling of the tidal force, its resultant deformation of the moon's interior, and the release of this stress as heat \citep{Dobos2015,Forgan2016a}.

In this letter, we compute the habitable zone for Earthlike exomoons orbiting Kepler-1625b assuming \citet{Teachey2018}'s derived parameters, and our exomoon climate models \citep{Forgan_moon1,Forgan2014a,Forgan2016a}.  We briefly describe the three versions of the climate model used to compute exomoon habitability in section \ref{sec:methods}, discuss the resulting habitable zones in section \ref{sec:results}, and conclude in section \ref{sec:conclusions}

\section{Methods}
\label{sec:methods}

\subsection{Simulation Setup}

\noindent We fix the stellar and planetary parameters, in accordance with \citet{Teachey2018} as follows.  The star mass $M_*= 1.04 \msol$.  The planet's mass $M_p=10\mjup$, and has radius $R_p = 1.015 \rjup$.  The planet's orbital semimajor axis and eccentricity are also fixed at $a_p=0.98$ and $e_p=0$ respectively.  This fixes the planet's Hill radius

\begin{equation}
R_{H,p} = a_p \left(\frac{M_p}{3M_*}\right)^{1/3} =0.1422\, \rm{AU} \approx 293.7 \,R_p
\end{equation}

We assume that the moon is Earthlike ($M_s = 1\mearth$, $R_s = 1\rearth$).  We further assume that the planet resides at the barycentre of the moon-planet system, which is satisfactory given that $\epsilon <<1$.  The inclination of the planet relative to the stellar equator, $i_p=0$ (i.e. the planet orbits in the $x-y$ plane).  

The inclination of the moon relative to the planet's equator, i.e. the inclination of the moon relative to the $x-y$ plane, $i_m$, is also zero (unless otherwise stated).  The orbital longitudes of the planet and moon are defined such that $\phi_{p}=\phi_{m}=0$ corresponds to the x-axis.  We also assume that the moon's obliquity has been efficiently damped by tidal evolution, and we therefore set it to zero.

We allow the moon's semimajor axis and eccentricity ($a_m$ and $e_m$ respectively) to vary.  For each model we run 500 simulations, where the lunar semimajor axis takes a range of values: $a_m=(0.05,0.3) R_{H,p}$, and eccentricities range from $e_m=(0,0.08)$. Note that  the exomoon candidate Kepler-1625b-i has an estimated orbital separation of $a_{m,i} = 0.153 R_{H,p}$.

\subsection{Latitudinal Energy Balance Modelling}

We compute the habitability of Earthlike exomoons using the Latitudinal Energy Balance Model (LEBM) approach \citep{North1979,North1983}.

The core equation of the LEBM is a diffusion equation, solved over latitude $\lambda \in (-90\degrees,90\degrees)$. In practice, we solve the equation using the convenience variable $x \equiv \sin \lambda$:

\begin{equation} 
C \frac{\partial T}{\partial t} - \frac{\partial }{\partial x}\left(D(1-x^2)\frac{\partial T}{\partial x}\right) = (S+S_p)\left[1-A(T)\right] + \zeta - I(T), \label{eq:LEBM}
\end{equation}

where $T=T(x,t)$ is the temperature at time $t$, and the boundary condition $\frac{dT}{dx}=0$ at the poles.  

$C$ is the atmospheric heat capacity, the diffusion coefficient $D$ controls latitudinal heat redistribution, $S$ and $S_p$ are the stellar and planetary illumination respectively, $\zeta$ is the surface heating generated by tides in the moon's interior, $I$ is the atmospheric infrared cooling and $A$ is the albedo.   

\begin{table*}
\caption{The 3 model runs in this paper, \label{tab:run-params}}
\begin{tabular}{cccccc}
\hline
Run Name & Infrared Cooling ($I$) & Planetary Illumination? & Carbonate-Silicate Cycle? & $i_m$ ($^\circ$) & $L$ \\
\hline
CN0 & Equation \ref{eq:simple_I} & No & No & 0 &  1.16 \\
CN45 & Equation \ref{eq:simple_I} & No & No & 45 &1.16 \\
CNL & Equation \ref{eq:simple_I} & No & No & 0 & 2.68 \\
IL & Equation \ref{eq:simple_I} & Yes & No & 0 & 1.16 \\
CS & Equation \ref{eq:WK97_I} & Yes & Yes & 0 & 1.16 \\
\hline
\end{tabular}
\end{table*}

We produce five measures of the habitable zone, using three different versions of the climate model, as described in Table \ref{tab:run-params}.  Essentially, these represent increasing realism for the energy balance model, as we add in the effect of planetary illumination, and the carbonate-silicate cycle.  They also consider the current luminosity of Kepler-1625 as it evolves off the main sequence, as well as its previous luminosity while on the main sequence.

\subsubsection{The Control Runs (CN)}

In the first set of three model runs (corresponding to the model used in \citealt{Forgan_moon1}), we use the following prescriptions.  

The atmospheric heat capacity $C$ depends on what fraction of the moon's surface is ocean, $f_{ocean}=0.7$, what fraction is land $f_{land}=1.0-f_{ocean}$, and what fraction of the ocean is frozen $f_{ice}$:

\begin{equation} 
C = f_{land}C_{land} + f_{ocean}\left[(1-f_{ice})C_{ocean} + f_{ice} C_{ice}\right]. 
\end{equation}

\noindent The heat capacities of land, ocean and ice covered areas are 

\begin{equation} 
C_{land} = 5.25 \times 10^9  \rm{erg \, cm^{-2} K^{-1}},
\end{equation}

\begin{equation} C_{ocean} = 40.0C_{land},\end{equation}
\begin{equation} C_{ice} = \begin{cases}
9.2C_{land} &  263 K <T < 273 K \\
2C_{land} &  T<263 K \\
0.0 & T> 273 K.\\
\end{cases}. 
\end{equation}

\noindent These parameters assume a wind-mixed ocean layer of 50m \citep{Williams1997a}.  Increasing the assumed depth of this layer would increase $C_{ocean}$ (see e.g. \citealt{North1983} for details).  The albedo function is

\begin{equation} 
A(T) = 0.525 - 0.245 \tanh \left[\frac{T-268\, \mathrm K}{5\, \mathrm K} \right]. 
\end{equation}

\noindent This produces a rapid shift from low albedo ($\sim 0.3$) to high albedo ($\sim 0.75$) as the temperature drops below the freezing point of water, producing highly reflective ice sheets.  Figure 1 of \citet{Spiegel_et_al_08} demonstrates how this shift in albedo affects the potential for global energy balance, and that for planets in circular orbits, two stable climate solutions arise, one ice-free, and one ice-covered.  Spiegel et al also show that such a function is sufficient to reproduce the annual mean latitudinal temperature distribution on the Earth.

The diffusion constant $D$ is calibrated such that a fiducial Earth-Sun climate system reproduces Earth's observed latitudinal temperature see e.g. \citealt{North1981, Spiegel_et_al_08}).  Planets that rotate rapidly experience inhibited latitudinal heat transport, due to Coriolis forces truncating the effects of Hadley circulation (cf \citealt{Farrell1990, Williams1997a}).  The partial pressure of $\cotwo$ also plays a role.  We follow \citet{Williams1997a} by scaling $D$ according to:

\begin{equation} 
D=5.394 \times 10^2 \left(\frac{\omega_d}{\omega_{d,\oplus}}\right)^{-2} \left(\frac{P_{\cotwo}}{P_{\cotwo,\oplus}}\right),\label{eq:D}
\end{equation}

\noindent where $\omega_d$ is the rotational angular velocity of the planet, and $\omega_{d,\oplus}$ is the rotational angular velocity of the Earth, and $P_{\cotwo,\oplus}=3.3 \times 10^{-4}  \,\rm{bar}$. In this run, the partial pressure of $\cotwo$ is fixed: $P_{\cotwo} = P_{\cotwo,\oplus}$.

The stellar insolation flux $S$ is a function of both season and latitude.  At any instant, the bolometric flux received at a given latitude at an orbital distance $r$ is

\begin{equation}
S = q_0 \left(\frac{M}{\msol}\right)^4\cos Z \left(\frac{1 AU}{r}\right)^2,
\end{equation}

\noindent where $q_0$ is the bolometric flux received from a $1\msol$ star at a distance of 1 AU, and we have assumed a standard main sequence luminosity relation.  $Z$ is the zenith angle:

\begin{equation} 
q_0 = 1.36\times 10^6\left(\frac{M_*}{\msol}\right)^4 \mathrm{erg \,s^{-1}\, cm^{-2}} 
\end{equation}

\begin{equation} 
\cos Z = \mu = \sin \lambda \sin \delta + \cos \lambda \cos \delta \cos h. 
\end{equation} 

\noindent $\delta$ is the solar declination, and $h$ is the solar hour angle.  As stated previously, we set the moon's obliquity $\delta_0$ to zero. The solar declination is calculated as:

\begin{equation} 
\sin \delta = -\sin \delta_0 \cos(\phi_{*m}-\phi_{peri,m}-\phi_a), 
\end{equation}

\noindent where $\phi_{*m}$ is the current orbital longitude of the moon \emph{relative to the star}, $\phi_{peri,m}$ is the longitude of periastron, and $\phi_a$ is the longitude of winter solstice, relative to the longitude of periastron.   We set $\phi_{peri,m}=\phi_a=0$ for simplicity. 

We must diurnally average the solar flux:

\begin{equation} 
S = q_0 \bar{\mu}. 
\end{equation}

\noindent This means we must first integrate $\mu$ over the sunlit part of the day, i.e. $h=[-H, +H]$, where $H$ is the radian half-day length at a given latitude.  Multiplying by the factor $H/\pi$ (as $H=\pi$ if a latitude is illuminated for a full rotation) gives the total diurnal insolation as

\begin{equation} 
S = q_0 \left(\frac{H}{\pi}\right) \bar{\mu} = \frac{q_0}{\pi} \left(H \sin \lambda \sin \delta + \cos \lambda \cos \delta \sin H\right). \label{eq:insol}
\end{equation}

\noindent The radian half day length is calculated as

\begin{equation} 
\cos H = -\tan \lambda \tan \delta. 
\end{equation}

\noindent We allow for eclipses of the moon by the planet (where the insolation $S=0$). We detect an eclipse by computing the angle $\alpha$ between the vector connecting the moon and planet, $\mathbf{s}$, and the vector connecting the moon and the star $\mathbf{s}_{*}$:

\begin{equation}
\cos \alpha = \mathbf{\hat{s}.\hat{s}_*} 
\end{equation}

\noindent It is straightforward to show that an eclipse is in progress if

\begin{equation}
\left|s_*\right| \sin \alpha < R_p
\end{equation}

\noindent We do not model the eclipse ingress and egress, and instead simply set $S$ to zero at any point during an eclipse.  A typical eclipse duration in these runs is approximately 6 hours (for an exomoon in a circular orbit around Kepler-1625b at $a_m=0.1 \,\rm{R_H}$). Our simulation timestep includes a condition to ensure that any eclipse must be resolved by at least ten simulation timesteps.

We use the following infrared cooling function:

\begin{equation} 
I(T) = \frac{\sigma_{SB}T^4}{1 +0.75 \tau_{IR}(T)}, \label{eq:simple_I}
\end{equation}

\noindent where the optical depth of the atmosphere 

\begin{equation} 
\tau_{IR}(T) = 0.79\left(\frac{T}{273\,\mathrm{K}}\right)^3. 
\end{equation}

\noindent Tidal heating is calculated by assuming the tidal heating per unit area is \citep{Peale1980,Scharf2006}:

\begin{equation} 
\left(\frac{dE}{dt}\right)_{tidal} = \frac{21}{38}\frac{\rho^2_m R^{5}_m e^2_m}{\Gamma Q}\left(\frac{GM_p}{a^3_m}\right)^{5/2} 
\end{equation}

\noindent where $\Gamma$ is the moon's elastic rigidity (which we assume to be uniform throughout the body), $R_m$ is the moon's radius, $\rho_m$ is the moon's density, $M_p$ is the planet mass, $a_m$ and $e_m$ are the moon's orbital semi-major axis and eccentricity (relative to the planet), and $Q$ is the moon's tidal dissipation parameter.  We assume terrestrial values for these parameters, hence $Q=100$, $\Gamma=10^{11} \,\mathrm{dyne \, cm^{-2}}$ (appropriate for silicate rock) and $\rho_m=5 \, \mathrm{g \, cm^{-3}}$.

We assume that this heating occurs uniformly across the moon's surface, which is a large approximation but a necessity of one-dimensional LEBM models.

The planetary illumination $S_p$ is set to zero for all three control runs.  Two of our three runs consider the habitable zone for $i_m=0^\circ$ (CN0) and $i_m=45^\circ$ (CN45), assuming the stellar luminosity is set by main sequence relations, giving 

\begin{equation}
\frac{L}{\lsol}= \left(\frac{M_*}{\msol}\right)^4 = 1.16
\end{equation}

\noindent These runs consider the main sequence phase of Kepler-1625b.  However, as previously stated Kepler-1625b is now evolving off the main sequence.  Therefore, in the last of the control runs, we reset $i_m=0^\circ$ and replace the main sequence luminosity with an estimate of Kepler-1625b's current luminosity, which is calculated assuming a blackbody and an effective temperature of $T_{\rm eff} = 5,548$ K \citep{Mathur2017}.  This results in a much greater luminosity of $L = 2.68 \lsol$.

\subsubsection{Planetary Illumination (IL)}

In this run, we use the same inputs as CN, (with $i_m=0$ and main sequence luminosity) but now implement planetary illumination and eclipses of the moon by the planet.  We use the same prescriptions as \citet{Forgan2014a} (see also \citealt{Heller2013}).  Illumination adds a second insolation source to the system, dependent on both the starlight reflected by the planet, and on the planet's own thermal radiation. 

We assume the planet's orbit is not synchronous, and that the temperature of the planet $T_p$ is uniform across the entire surface. We also fix the planetary albedo $\alpha_p=0.3$.  The insolation due to the planet is therefore

\begin{equation}
S_p(t)=f_t(t)+f_r(t).
\end{equation}

\noindent The thermal flux $f_t$ as a function of latitude $\lambda$ is:

\begin{equation}
f_t(\lambda,t)= \frac{2R_p^2 \sigma_{SB}}{a_{m}^2} (\cos \lambda) T_{p}^4,
\end{equation}

\noindent And the reflected flux $f_r$ is:

\begin{equation}
f_r(\lambda,t)=\frac{2L_*}{4\pi r_{p*}^2} \frac{R_p^2 \pi \alpha_p}{a_{m}^2} \cos \lambda \Xi(t).
\end{equation}

\noindent Where $\Xi$ is the fraction of dayside visible from the lunar surface (see \citealt{Forgan2014a} for more details). Calculating the ratio $f_r/f_t$ for Kepler-1625b shows that thermal flux dominates the planetary illumination budget, with around 1\% of the total illumination being produced by reflected starlight.  We should therefore expect $S_p$ to be roughly constant over the moon's orbit (as $T_p$ is uniform).

\subsubsection{The Carbonate Silicate-Cycle (CS)}

In this final run, we now use a simple piecewise function to determine $P_{\cotwo}$ as a function of local temperature \citep{Spiegel2010}:

\begin{equation}
P_{\cotwo} = \begin{cases} 
10^{-2} \, \rm{bar} & T \leq 250 K \\
10^{-2 - (T-250)/27} \,\rm{bar} & 250K < T < 290K \\
P_{\cotwo, \oplus} & T \geq 290 K
\end{cases}
\end{equation}

\noindent Our prescription now allows $D$ to vary with latitude, depending on the local temperature.  This is not guaranteed to produce Hadley circulation (see e.g. \citealt{Vladilo2013} for details on how $D$ can be modified to achieve this).  As we allow partial pressure of $\cotwo$ to vary, we now replace equation \ref{eq:simple_I} with  \citet{Williams1997a}'s prescription for the cooling function, $I(T,P_{\cotwo})$:

\begin{multline}
I = 9.468980 -7.714727 \times 10^{-5} \beta - 2.794778T  \\
 - 3.244753 \times 10^{-3} \beta T -3.4547406 \times 10^{-4}\beta^2 \\
 + 2.212108 \times 10^{-2} T^2 + 2.229142 \times 10^{-3} \beta^2 T  \\
 + 3.088497 \times 10^{-5} \beta T^2 - 2.789815 \times 10^{-5} \beta^2 T^2 \\
 - 3.442973 \times 10^{-3} \beta^3 - 3.361939 \times 10^{-5} T^3 \\
 + 9.173169 \times 10^{-3} \beta^3 T - 7.775195 \times 10^{-5} \beta^3 T^2 \\
 - 1.679112 \times 10^{-7} \beta T^3 + 6.590999 \times 10^{-8} \beta^2 T^3 \\  
 +1.528125 \times 10^{-7} \beta^3 T^3 - 3.367567 \times 10^{-2} \beta^4 \\
 -1.631909 \times 10^{-4} \beta^4 T + 3.663871 \times 10^{-6} \beta^4 T^2 \\
 -9.255646 \times 10^{-9} \beta^4 T^3 \label{eq:WK97_I}
\end{multline}

\noindent where we have defined 

\begin{equation}
\beta = \log \left(\frac{P_{\cotwo}}{P_{\cotwo,\oplus}}\right).
\end{equation}

\subsection{Simulation Timestep}

\noindent The diffusion equation is solved using a simple explicit forward time, centre space finite difference algorithm.  A global timestep was adopted, with constraint

\begin{equation}
\delta t < \frac{\left(\Delta x\right)^2C}{2D(1-x^2)}.  
\end{equation}

\noindent This timestep constraint ensures that the first term on the left hand side of equation (\ref{eq:LEBM})  is always larger than the second term, preventing the diffusion term from setting up unphysical temperature gradients.  The parameters are diurnally averaged, i.e. a key assumption of the model is that the moons rotate sufficiently quickly relative to their orbital period around the primary insolation source.  This is broadly true, as the star is the principal insolation source, and the moon rotates relative to the star on timescales of a few days.

\subsubsection{Mapping the Exomoon Habitable Zone}

For each run, we simulate climate models over a range of lunar semimajor axes $a_m$ and eccentricities $e_m$, mapping out the habitable zone as a function of $(a_m,e_m)$ by classifying the resulting climate according to its habitability function $\xi$:

\begin{equation} 
\xi(\lambda,t) = \left\{
\begin{array}{l l }
1 & \quad \mbox{273 K $< T(\lambda,t) <$ 373 K} \\
0 & \quad \mbox{otherwise}. \\
\end{array} \right. 
\end{equation}

\noindent We average this over latitude to calculate the fraction of habitable surface at any timestep:

\begin{equation} 
\xi(t) = \frac{1}{2} \int_{-\pi/2}^{\pi/2}\xi(\lambda,t)\cos \lambda \, d\lambda. 
\end{equation}

\noindent Each simulation evolved until it reaches a steady or quasi-steady state, and the final ten years of climate data are used to produce a time-averaged value of $\xi(t)$, $\bar{\xi}$.  Along with the sample standard deviation, $\sigma_{\xi}$, we can classify each simulation as follows:

\begin{enumerate}
\item \emph{Habitable Moons} - these moons possess a time-averaged $\bar{\xi}>0.1$, and $\sigma_{\xi} < 0.1\bar{\xi}$, i.e. the fluctuation in habitable surface is less than 10\% of the mean.
\item \emph{Hot Moons} - these moons have average temperatures above 373 K across all seasons, and are therefore conventionally uninhabitable, and $\bar{\xi} <0.1$.
\item \emph{Snowball Moons} - these moons have undergone a snowball transition to a state where the entire moon is frozen, and are therefore conventionally uninhabitable\footnote{As with hot moons, we require $\bar{\xi}<0.1$ for the moon to be classified as a snowball, but given the nature of the snowball transition as it is modelled here, these worlds typically have $\bar{\xi}=0$.}.
\item \emph{Transient Moons} - these moons possess a time-averaged $\bar{\xi}>0.1$, and $\sigma_{\xi} > 0.1\bar{\xi}$, i.e. the fluctuation in habitable surface is greater than 10\% of the mean.
\end{enumerate}

\section{Results \& Discussion}
\label{sec:results}

\subsection{Control Runs}

\begin{figure*}
\begin{center}$
\begin{array}{cc}
\includegraphics[scale=0.45]{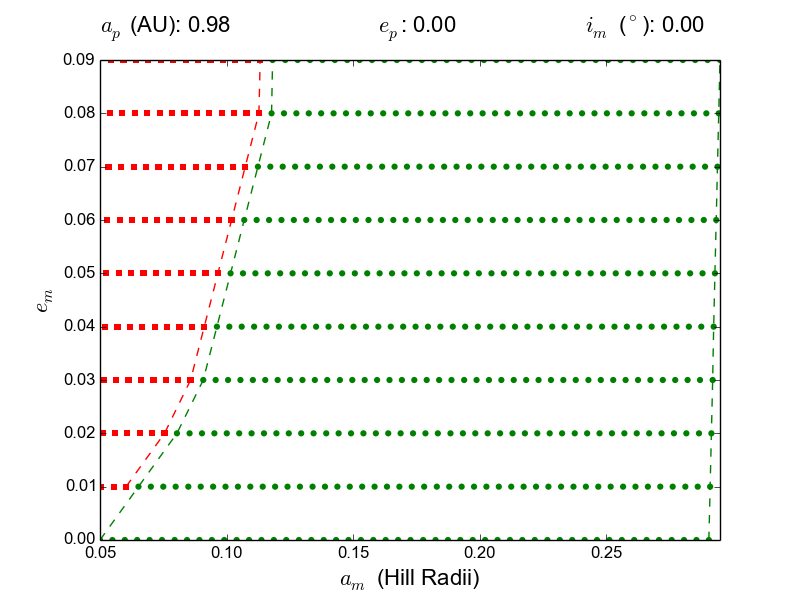} \\
\includegraphics[scale=0.45]{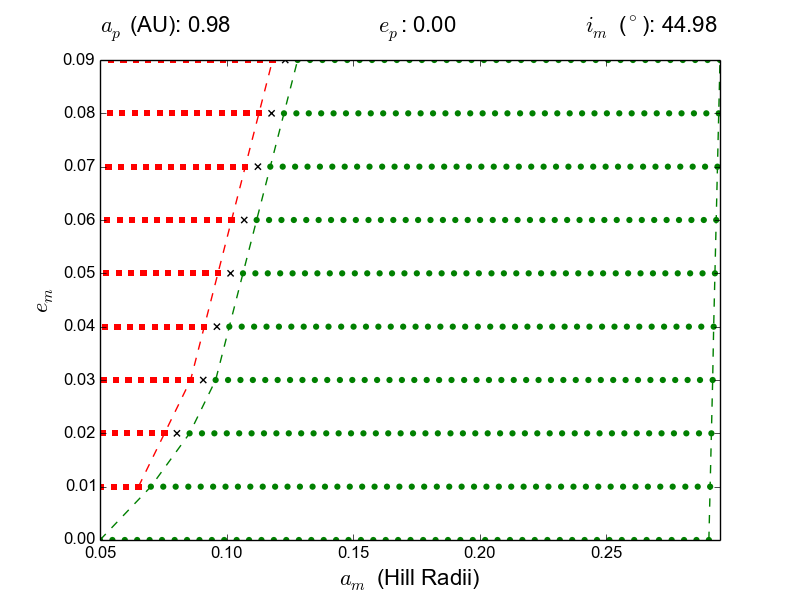}\\
\includegraphics[scale=0.45]{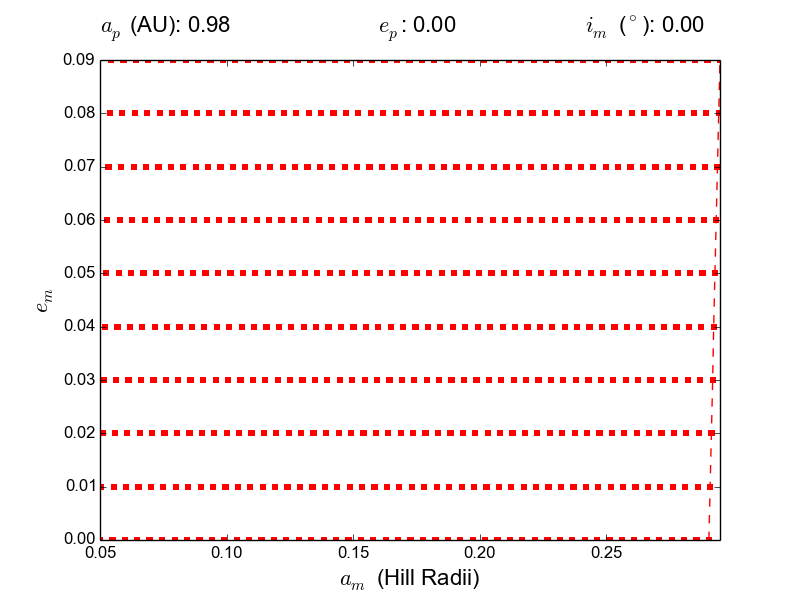} \\
\end{array}$
\end{center}
\caption{The habitable zone for exomoons orbiting Kepler-1625b, for the three control model runs CN.  Top: The habitable zone for $i_m=0^\circ$, assuming stellar luminosity consistent with the main sequence (CN0).  Middle: The same, but for $i_m=45^\circ$ (CN45).  Bottom: The habitable zone for $i_m=0$, using estimates of Kepler-1625b's current luminosity, rather than that derived from main sequence relations (CNL).  Note that the candidate exomoon Kepler-1625b-i orbits at 0.153 Hill radii.  If Kepler-1625b-i exists, then any other exomoon orbiting with $a_m <0.17$ Hill radii is likely to be dynamically unstable. \label{fig:CNresults}}
\end{figure*}

In Figure \ref{fig:CNresults}, we display results for the CN0, CN45 and CNL runs.  For CN0 and CN45, orbits with a semimajor axis above 0.1 $R_{H,p}$ are habitable for a wide range of eccentricities.  There is no outer circumplanetary habitable edge, despite all bodies rotating in the same plane, maximising the effect of eclipses. The effect of inclination results in some moons near the habitable zone inner boundary experiencing large temperature variations, but otherwise maintaining a habitable surface.

Notably, there are no habitable moons when Kepler-1625b's luminosity is increased from its main sequence value to its current estimated value.  Extra runs considering larger $a_m$ also fail to find any habitable solutions.

\subsection{Planetary Illumination and the Carbonate-Silicate Cycle}

\begin{figure*}
\begin{center}$
\begin{array}{cc}
\includegraphics[scale=0.45]{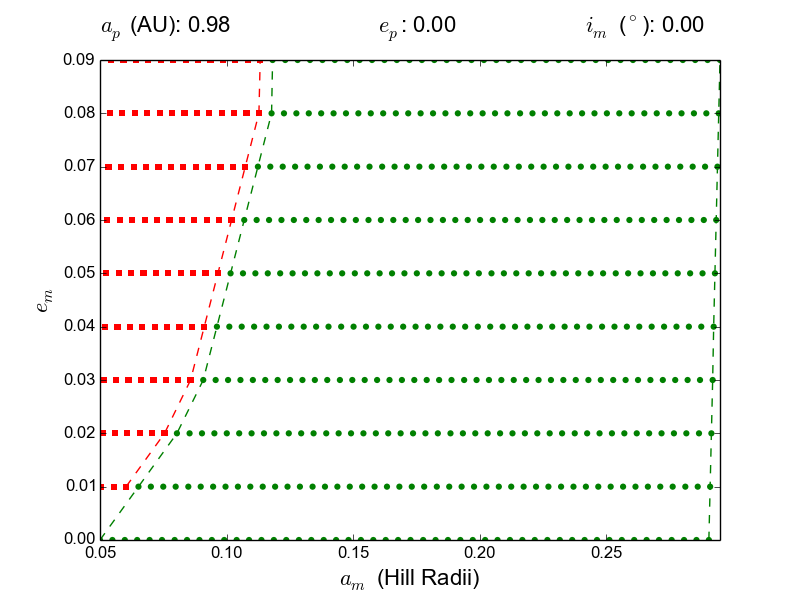}\\
\includegraphics[scale=0.45]{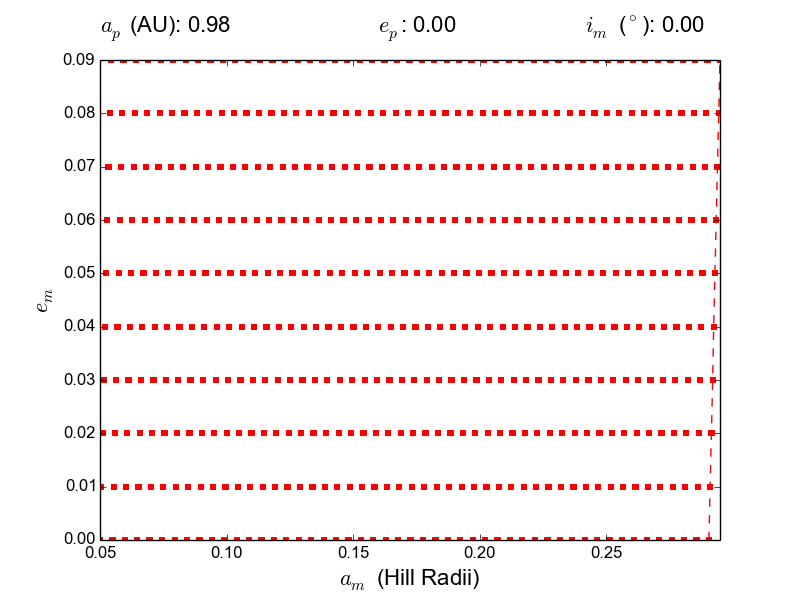} \\
\end{array}$
\end{center}
\caption{The habitable zone for exomoons orbiting Kepler-1625b, for model runs IL (top) and CS (bottom).  Note that the candidate exomoon Kepler-1625b-i orbits at 0.153 Hill radii.  If Kepler-1625b-i exists, then any other exomoon orbiting with $a_m <0.17$ Hill radii is likely to be dynamically unstable. \label{fig:results}}
\end{figure*}

\noindent In Figure \ref{fig:results} we show the results from the IL and CS runs (with $i_m=0$).  Planetary illumination makes little appreciable impact on the habitable zone, which is consistent with previous calculations \citep{Forgan2014a}.  Ideally, as the planetary illumination is typically in the infrared, it should be subject to a different albedo than the stellar illumination \cite[cf][]{Heller2015}.  Experimenting with different albedos for $S_p$ make little appreciable difference to the results.

When $\cotwo$ partial pressure is allowed to vary (run CS), a runaway greenhouse takes effect in all runs, preventing the moon from sustaining a habitable surface.  Further runs have demonstrated that this is independent of tidal heating, and that extending the parameter sweep to larger $a_m$ does not yield a habitable solution.  

\subsection{Orbital stability of moons due to Kepler-1625b-i}

Moons remain orbitally stable for semimajor axes $a_m < 0.3-0.5 R_{H,p}$  \citep{Domingos2006}, and we can therefore expect all moons simulated in this work to remain on stable orbits for long timescales, provided they do not orbit too close to Kepler-1625b-i (if it exists).  Taking \citet{Teachey2018}'s estimate of Kepler-1625b-i as approximately Neptune-mass $\approx 17\mearth$, then we should expect an inner orbital stability limit defined roughly by the Hill radius of the satellite candidate:

\begin{equation}
R_{H,i} = a_{m,i} \left(\frac{M_s}{3 M_p}\right)^{1/3} = 0.019 R_{H,p} 
\end{equation}

\noindent Where $a_{m,i}=0.153 R_{H,p}$.  Any satellite which orbits within a few mutual Hill radii of the candidate will be subject to dynamical instability.  We can therefore expect that any Earthlike exomoon orbit will only be stable provided $a_m \gtrapprox 0.17-0.2 R_{H,p}$.  This still leaves a large range of orbits for Earthlike exomoons that are both habitable (depending on the input climate model) and dynamically stable.

We note that while many Earthlike exomoons can orbit stably in the Kepler-1625b system despite the presence of Kepler-1625b-i, we have not computed the cyclic variations in orbital eccentricity we might expect as a Neptune-sized body exerts its gravitational field on a neighbour.  The large mass and inclination of Kepler-1625b-i is likely to drive strong gravitational perturbations.  If the eccentricity of a body is driven too high, tidal heating could rapidly render an Earthlike exomoon uninhabitable.  We are currently running climate calculations using OBERON \citep{Forgan2016f} that includes the gravitational interaction between bodies to explore this further. 

Clouds are not explicitly considered in this model (although they are implicitly accounted for in run CS).  Clouds can modify both the albedo and optical depth of the system significantly.  Also, we assume that both stellar and planetary flux are governed by the same lunar albedo, which in truth is not likely to be the case (see e.g. \citealt{Heller2013b}).  Planetary flux at infrared wavelengths is more easily absorbed by ice sheets and produces more efficient melting \citep[see e.g.][]{Shields2013}.  However, our simulations do not produce large quantities of ice in any of the model runs.  It is possible that more efficient absorption of IR radiation might move the inner habitable zone further outwards in $a_m$, but this can be equally offset by cloud cover.

In several runs, we find that an Earthlike exomoon at the calculated location of Kepler-1625b-i would possess a habitable climate.  Given that Kepler-1625b-i appears to be a relatively massive exomoon, we can consider the possibility that this exomoon candidate could have its own satellite (a ``moon-moon'').   If we assume an Earthlike satellite of the exomoon candidate, we can compute the minimum and maximum permitted orbital semimajor axies.  The Hill radius of the satellite candidate is approximately $5 R_p$, giving an outer stability limit of approximately $2.5 R_p \approx 5.7 R_s$ \citep{Domingos2006}.  The inner stability limit is defined by tidal disruption.  Setting $M_{ss}=1\mearth$ to be the mass of the moon-moon, and $R_{ss}=1\rearth$, we can compute the Roche radius:

\begin{equation}
R_{\rm Roche, ss} = R_{ss}\left(\frac{M_s}{M_{ss}}\right)^{1/3} = 0.5 R_s
\end{equation}

\noindent There therefore exists a reasonable range of distances, between say $[2,5.7] R_s$, at which an Earthlike moon-moon could orbit the exomoon candidate.  If this moon-moon was Earthlike, our models suggest it would have been habitable during the star's main sequence phase, and its climate would be quite analogous to the climates of planets in binary systems (cf \citealt{Forgan2012,Kaltenegger2013}).

\section{Conclusions}
\label{sec:conclusions}

\noindent We have applied several different exomoon climate models to the Kepler-1625b system, to investigate the morphology of the exomoon habitable zone around this Jupiter-radius object.  

We find that for a range of assumptions, Earthlike exomoons were likely to have been habitable while Kepler-1625 occupied the main sequence, for a wide range of orbital semimajor axes and eccentricities.  These exomoons would remain orbitally stable even if the exomoon candidate Kepler-1625b-i is indeed present.  However, as Kepler-1625 evolved off the main sequence, the luminosity increased to levels that generally destroy the habitability of any Earthlike exomoons possibly present. 

Exomoon detection remains a challenging observational endeavour - as such, the ability to determine the habitability of detected exomoons is an even greater challenge.  The four classes of exomoon climate displayed in this work will be extremely difficult to distinguish between.  The best approaches will require some form of spectroscopic data to assess atmospheric composition and thermal state. This could be obtained if the moon is sufficiently bright compared to the planet at wavelengths of interest.  Proposed techniques for obtaining exomoon spectral data involve spectroastrometry \citep{Agol2015}, oscillations in the combined exoplanet-exomoon phase curve \citep{Forgan2017b}, or radial velocity measurements of the exoplanet using high dispersion spectroscopy (Brogi \& Forgan, in prep.).

If Kepler-1625b-i is real, it is likely massive enough to possesses its own satellite (a ``moon-moon'').  If the said moon-moon was Earthlike, it could have resided in a moon-moon habitable zone during Kepler-1625's main sequence phase. The morphology of moon-moon habitable zones are not yet explored, but will share similarities with that of S-type binary star systems \citep{Kaltenegger2013,Cuntz2014,Forgan2016c}.  

As an aside, we should note that one dimensional calculations are now giving way to full 3D global circulation models of exomoon atmospheres \citep{Haqq-Misra2018a}.  The 3D aspect of exomoon climates is crucial, as planetary illumination heats the top of the atmosphere, and tidal effects heat the surface, resulting in complex heat redistribution patterns.  For example, planetary illumination amplifies warming at the moon's poles, an effect not seen in 1D calculations.  We look forward to further 3D studies of exomoon atmospheres that explore the habitable zone's orbital parameters as defined by studies such as this work.

We suggest that future studies of habitability of rocky worlds should continue to explore what have until now been considered rather unusual regimes in parameter space, as it seems likely the Universe will continue to deliver surprising configurations for celestial objects.

\section*{Acknowledgments}

The author gratefully acknowledges support from the ECOGAL project, grant agreement 291227, funded by the European Research Council under ERC-2011-ADG.  The author warmly thanks the reviewer for their comments and suggestions. This  research  has  made  use  of  NASA's Astrophysics Data System Bibliographic Services.

\bibliographystyle{mnras} % (must include a bibliography style)
\bibliography{K1625b_habzone}

\label{lastpage}

\end{document}